\documentclass[aps,amssymb,article,pra,twocolumn,showpacs,floatfix]{revtex4}

\usepackage{epsf}
\usepackage{amsmath}
\usepackage{graphicx}
\begin{document}
         
\title{Anomalous electron states}

\author{Boris I. Ivlev}

\affiliation{Instituto de F\'{\i}sica, Universidad Aut\'onoma de San Luis Potos\'{\i},\\ 
San Luis Potos\'{\i}, 78000 Mexico}

\begin{abstract}

In experiments \cite{IG10,IG11} on irradiation of metal surfaces by ions of $keV$ energy, the emission of X-ray laser beams from the metal was observed not only during the irradiation but also 20 
hours after it was switched off (from the ``dead'' sample). In contrast to an usual laser, the emitted collimated X-ray beams were of continuous frequency. In this paper the mechanism of that 
phenomenon is proposed. Subatomic electron states are formed inside the metal. These states are associated with anomalous well within the subatomically small ($10^{-11}cm$) region. Anomalous well 
is formed by the local reduction (of $MeV$ scale) in that region of the vacuum energy of the mass-generating field. States in anomalous well are long-living which results in population inversion 
and the subsequent laser generation observed. The energy of emitted X-ray beams are due to the conversion of the vacuum energy of the mass-generating field (X-ray laser beams from vacuum).

\end{abstract} \vskip 1.0cm


\maketitle

\section{INTRODUCTION}
\label{intr}
The field of surface irradiation by ions is well studied \cite{IR1,IR2,IR3,IR4,IR5,IR6,IR7,IR8,IR9}. An external ion beam may result in photon emission from the surface of a solid 
\cite{IG5,IG6,IG7,IG8,IG9}. Underlying mechanisms of this phenomenon are successfully described in terms of known effects in condensed matter physics. Nobody could expect that in this field 
there can be something which may turn the mind from the common track \cite{IG10,IG11}. 

In experiments \cite{IG10,IG11} on irradiation of metals by $keV$ ions a lot of nuclear transmutations, required $MeV$ energies, were revealed. Moreover collimated X-ray bursts in the $keV$ range
were registered during the ion irradiation and even 20 hours after switching it off. Therefore long-living states existed to produce the laser effect. This means that the usual purse quarter, 
irradiated yesterday evening by $keV$ ions, can emit X-ray laser bursts. In contrast to a conventional monochromatic laser radiation, the observed one was of continuous frequency. The experiments 
\cite{IG10,IG11} were repeatedly performed for years and could be reproduced any time on demand. 

The underlying mechanism was a mystery since nuclear processes were not responsible for the $MeV$ energies involved. This is because the $keV$ energies of irradiating ions are not sufficient, as 
known, to directly cause nuclear processes due to high Coulomb barriers around nuclei. Also the existence of long-living states in the $keV$ range is incompatible with nuclear spectroscopy 
\cite{IG10}. This points to a high-energy processes resulted from a not nuclear source. 

High energies of the $MeV$, or even $keV$, scale relate to subatomic processes. The situation with the high energy emission in condensed matter physics looks paradoxical. Electrons, interacting 
solely with crystal lattice, including the static field and phonons, cannot relate to $MeV$ energies and therefore to a subatomic scale. But what can happen to the scale if to supplement that 
interaction by electromagnetic one? 

To answer that question let us look at the shift of electron energy levels by its interaction with photons (the Lamb shift \cite{LANDAU}). Under this interaction the electron ``vibrates'' with
the mean displacement $\langle\vec u\rangle=0$ and the mean squared displacement $r^{2}_{T}=\langle u^2\rangle$ where $r_T\sim 10^{-11}cm$ \cite{MIGDAL,BOY}. This is fluctuation spreading in 
addition to the usual quantum mechanical uncertainty. So the electron becomes less ``heavy''. In this language, ``vibrating'' electron probes various parts of the static potential and therefore 
changes its energy. In the usual case this energy variation is small since the spatial distribution of the electron in the static potential essentially exceeds $r_T$. Note that for the electron, 
interacting solely with photons (no mean-field forces on electrons), $r_T=\infty$ \cite{MIGDAL}. 

The aforementioned remarks provide a hint for the origin of a short scale in the electron-photon system. Without the interaction with photons the electron wave function can be singular on the 
certain line. This situation is not physical since this singularity, in the term $-\hbar^2\nabla^2/2m$, is not supported by a singular potential well along the line. 

With the interaction with photons, due to electron ``vibrations'' the singularity on the line gets smeared out into the thin thread of the radius $\sim r_T$. Within the thread the term 
$-\hbar^2\nabla^2/2m$ goes over into $\hbar^2/mr^{2}_{T}$. As shown in the paper, that large kinetic energy is supported by counter-terms that can be interpreted as anomalous well along the thread. 
Those terms are formed by the variation in space of the energy of the mass-generating field. In the case of electron this is the Higgs field. The processes, generic with electron mass generation 
and localized within the electron radius, are distributed due to electron ''vibrations`` caused by photons. 

Inside the thread anomalous well depth $\hbar c/r_T\sim mc^2\sqrt{\hbar c/e^2}\sim 1MeV$ is singular with respect to $e^2/\hbar c$. Electron states inside anomalous well are exact (non-decaying) 
and with the continuous spectrum. This is compatible with population inversion leading to the unexpected X-ray laser generation with the continuous spectrum. Transitions down in energy in 
anomalous $MeV$ wells produce $MeV$ quanta which can cause nuclear transmutation observed in \cite{IG10,IG11}. Those processes, involving the $MeV$ energy range, are not due to nuclear fusion, as 
supposed in Refs.~\cite{IG10,IG11}, but of the electron origin.

The emitted energy comes from vacuum ("energy from nothing``) of the mass-generating field. The steady conversion of $keV$ energy of irradiating ions into $MeV$ energy of generated quanta (steady 
extraction of vacuum energy) can be significant for applications. 

Anomalous electron states are responsible not only for phenomena observed in experiments \cite{IG10,IG11}. The recent observations of anomalous oscillations of magnetoresistance in superconductors 
\cite{MIL} provide another mysterious example generic with \cite{IG10,IG11}. The paradoxical universality of the oscillation phenomenon (in particular, material independence) can be explained 
solely by a subatomic mechanism. The proposed subatomic mechanism, based on electron anomalous states \cite{IVLEV1}, provides an excellent qualitative explanation of the experimental results
\cite{MIL}.

In Sec.~\ref{kar} the experiments \cite{IG10,IG11} are analyzed in the extended manner. In Secs.~\ref{well} and \ref{qed} the connection to quantum electrodynamics is studied. In Sec.~\ref{anom} 
the formation of anomalous electron states is described. In Sec.~\ref{exp} the link from experiments \cite{IG10,IG11} to anomalous electron states is analyzed. Sec.~\ref{disc} is the the general 
view on the problem.
\section{UNUSUAL X-RAY LASER BEAMS FROM SOLIDS}
\label{kar}
\subsection{Description of the experiment}
\label{karA}
In papers \cite{IG10,IG11} a photon emission from various metals, under the action of glow discharge, was studied. The glow discharge provides an 
ion beam on the metal surface. The basic instrumentation is the glow discharge chamber with the metallic cathode of $1cm^2$ area. 
Various metals were used, Al, Sc, V, Ti, Ni, Zr, Mo, Pd, Ta, and W. Under the pressure of $(3-10)Torr$ the chamber is filed out by one of the gases 
${\rm D}_2$, ${\rm H}_2$, He, Kr, Ar, and Xe. The current can be chosen as $300mA$ and the glow discharge voltage as $(1000-4500)V$. Approximately 
$20cm$ from the cathode the bent mica crystal X-ray spectrometer is placed as shown in Figs.~\ref{fig1} and \ref{fig2}. The size of that 
spectrometer is of a few centimeters.

At discharge voltage of $(1-2)keV$ X-ray emission (up to $10keV$) from the metal cathode was registered. Both diffuse and collimated X-ray bursts of 
the duration of $20\mu s$ were registered approximately every $50\mu s$ during $0.1s$ after stopping the discharge (post-irradiation emission).

Moreover, some collimated X-ray bursts have been seen up to 20 hours after switching off the discharge voltage. As known, an emission of separate 
photons by radioactive isotopes from the cathode material is easy understandable. But in contrast, here one deals with strongly collimated X-ray 
laser bursts. So it was the laser emission from ``dead'' sample, namely, which was acted by nothing during 20 hours. 

The essential point is that experiments \cite{IG10,IG11} were repeatedly performed for years and could be reproduced any time on demand. Indeed, the array of macroscopic laser bursts unlikely 
is an artifact. In addition, multiple element transmutations were observed in \cite{IG10,IG11}.

The problem was to study the spectrum of short ($20\mu$s) pulses. It was impossibility to use the standard technique of slow adjusted Bragg 
spectrometer. Therefore the case of short pulses required more efforts. 
\subsection{What appears in the experiment}
\label{karB}
\begin{figure}
\includegraphics[width=5cm]{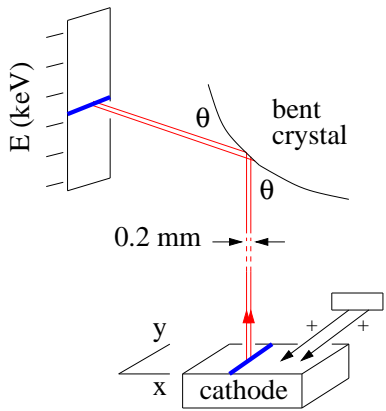}
\caption{\label{fig1}Collimated monoenergetic photon beams of the energy $E$ are emitted from points on the cathode surface with the same 
coordinate $x$. Each beam is reflected by the bent crystal spectrometer which is a cylinder along the $y$ axis. The image on the X-ray film (left 
in the figure) appears solely under the condition (\ref{1}) between $E$ and $x$. The area of the cathode is $9mm\times 9mm$. Its distance to the 
spectrometer is $20cm$. The radius of that spectrometer is $2.5cm$. The distance from it to the X-ray film is on the order of $1cm$.}
\end{figure}

The schematic illustration of the experimental setup is in Fig.~\ref{fig1}. Collimated laser beams are reflected from the bent crystal spectrometer 
according to Bragg's condition $\lambda(nm)=2.0\sin\theta$ for the mica crystal used. Accounting for the relation for photon energy 
$E(keV)=1.235/\lambda(nm)$, one can obtain the dependence of reflection angle $\theta$ in Fig.~\ref{fig1} on photon energy
\begin{equation}
\label{1}
\sin\theta(x)=\frac{0.617}{E(keV)}\,.
\end{equation}

The $x$ dependence of the angle $\theta(x)$ is determined by geometrical conditions of the setup in Fig.~\ref{fig1}. If the narrow beam is  
monoenergetic with the energy $E$, corresponding emission points should be of the certain coordinate $x$, given by (\ref{1}), to result in an
image on the X-ray film as in Fig.~\ref{fig1}. 

\begin{figure}
\includegraphics[width=5cm]{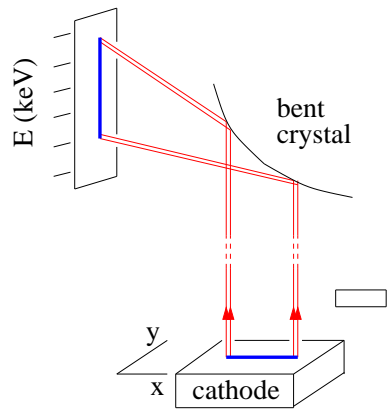}
\caption{\label{fig2}Emission points, of X-rays with the continuous spectrum, are distributed in the $x$ direction. Images on the X-ray film are
related to various energies.}
\end{figure}

When the narrow beam contains a continuous photon spectrum then for each emission point, with the coordinate $x$, the certain energy $E(x)$ 
(\ref{1}) exists in the spectrum to provide the related image point on the X-ray film as in Fig.~\ref{fig2}. 

\begin{figure}
\includegraphics[width=4.5cm]{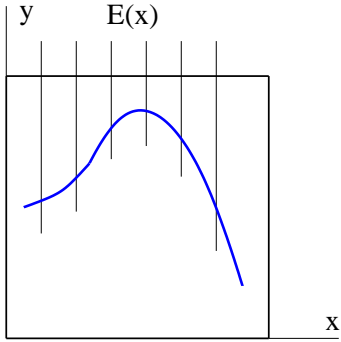}
\caption{\label{fig3}Example of a track of the emission point on the cathode surface (restricted by the black frame). Each point on the track, with 
the coordinate $x$, is produced by the particular energy $E(x)$ from the total continuous X-ray spectrum.}
\end{figure}

Tracks, obtained in \cite{IG10,IG11} on the X-ray film, are of $0.2mm$ width within $1cm$ length. This occurs since the emission point of the 
narrow beam, with a continuous spectrum, moves on the cathode surface during the $20\mu$s emission process. Without a motion it would be a point 
on the X-ray film. The example of the line track in the x-direction is sketched in Fig.~\ref{fig2}. The track of the emission point of more general 
form on the cathode surface is in Fig.~\ref{fig3}. This track can be obtained from one on the X-ray film by accounting for the geometric relation 
$\theta(x)$. 

\begin{figure}
\includegraphics[width=8cm]{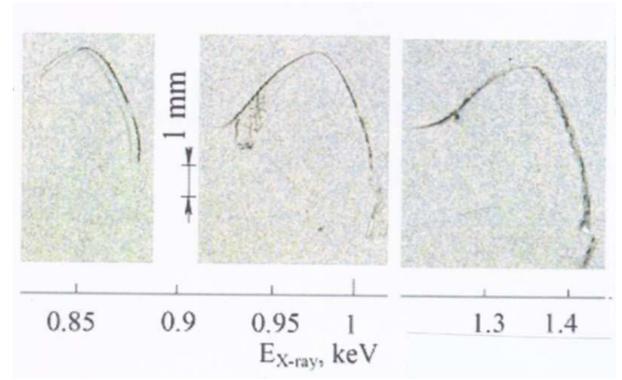}
\caption{\label{fig4}Observed tracks of the image point on the X-ray film gauged in units of energy as in Figs.~\ref{fig1} and \ref{fig2} 
\cite{IG10,IG11}. Pd cathode was used in the hydrogen gas.}
\end{figure}

Examples of obtained images on the X-ray film are shown in Fig.~\ref{fig4} \cite{IG10,IG11}. Mapping of these images on the cathode surface 
qualitatively remind the curve in Fig.~\ref{fig3}. 

Angular uncertainty of emitted bursts also can result in a curve (instead of point) track on the X-ray film. But this effect is small since, due to 
the geometry, the related uncertainty $\delta\theta\sim 0.9cm/20cm$ can lead to a track line of $1mm$ length. In experiments, close to the bent 
crystal, there is the the slit of $6mm$ wide (not shown in Figs.~\ref{fig1} and \ref{fig2}) which does not allow substantial angular uncertainty. 
In experiments \cite{IG10,IG11} the cathode-spectrometer distance was varied a few times but the burst were remained collimated.

In summary, in experiments \cite{IG10,IG11} (i) narrow collimated bursts were emitted from the cathode surface, (ii) the emission point moved on 
the cathode surface, and (iii) the bursts were of the continuous energy spectrum. 
\subsection{X-ray laser versus other radiation phenomena}
\label{karC}
The question is why the emitted burst is narrow and collimated. In principle, it could be a beam of the usual light emitted in the focus of a 
parabolic mirror and then reflected from it. But in experiments \cite{IG10,IG11} there were no conditions for that. 

In experiments \cite{IG10,IG11} the phenomenon of superradiance \cite{IG15} is also impossible since the emitted spectrum is continuous. Due to 
that there is no the certain singular transition which may multiply occur over the all entangled state. 

The only a reason for the emission of collimated beams is stimulated emission giving rise to laser effects.
\subsection{Looking for a mechanism}
\label{karD}
No one element 
\begin{itemize}
\item paradoxical X-ray laser generation despite the long-living states, resulting in population inversion, do not exist in nuclear spectroscopy, 
\item paradoxical continuous spectrum of the laser generation,
\item paradoxical generation of X-ray laser bursts from a ``dead'' sample during 20 hours after switching off the external source,
\item paradoxical observation of element transmutation required $MeV$ energies while only $keV$ was pumped in
\end{itemize}
of the experimental puzzle can be explained by a combination of known effects.

First, it is unclear how the energy inside the isolated and equilibrium solid is suddenly collected to get converted into the macroscopic laser 
burst. Second, even if this happens, a mechanism of creation of population inversion is also unclear since lifetimes in the $keV$ regime are short. 
Indeed, an excitation of nuclear degrees of freedom by $keV$ ions is not effective and nuclear lifetimes (no longer than $10^{-7}s$) are definitely 
less that $0.1s$ (moreover, than 20 hours). Lifetime of $keV$ electrons is also short. 

This means that the post-irradiation emission of $keV$ photons is not due to consuming of the energy stored 20 hours back. In contrast, the energy
for each burst is collected somehow before its generation.

Misinterpretation of experiments \cite{IG10,IG11} is possible by attributing the energy source to nuclear reactions. These reactions are impossible 
here since energies of phonons ($0.01eV$) and electrons ($1eV$) inside a solid are too low compared to $MeV$ range. It is not real to expect phonons 
in a solid to suddenly get collected into the $MeV$ energy. 

As we see, there is the paradoxical contradiction of the observed phenomena and known mechanisms. It happens that the different mechanism is responsible for unusual observations. See 
Sec.~\ref{anom}.
\section{ELECTRON IN THE WELL}
\label{well}
Suppose in the three-dimensional potential well $U(R)$ ($R^2=r^2+z^2$) the ground state energy of the electron is $E$ in the absence of the interaction with photons. Under this interaction the 
electron ``vibrates'' with displacements $\vec u$ \cite{MIGDAL}. The related mean displacement $\langle\vec u\rangle=0$ but the mean squared displacement $r^{2}_{T}=\langle u^2\rangle$ is finite. 
The effective potential can be estimated as 
\cite{MIGDAL}
\begin{equation}
\label{51}
\langle U(|\vec R-\vec u|)\rangle\simeq U(R)+\frac{\langle u^2\rangle}{6}\nabla^2U(R).
\end{equation}
The quantum mechanical perturbation, due to the second term in (\ref{51}), leads to the energy $E_{tot}$ which is shifted (the Lamb shift) with respect to $E$ \cite{LANDAU}. 
\begin{equation}
\label{52}
E_{tot}=E+\frac{\langle u^2\rangle}{6}\int\psi^*(\vec R)\nabla^2 U(\vec R)\psi(\vec R)d^3R.
\end{equation}
For hydrogen atom $U(R)=-e^2/R$, $\nabla^2U(\vec R)=4\pi^2e^2\delta(\vec R)$, and $|\psi(0)|^2=(me^2/\hbar^2)^3/\pi$. With the logarithmic accuracy for the ground state \cite{LANDAU}
\begin{equation}
\label{53}
E_{tot}=E+\frac{8mc^2}{3\pi}\left(\frac{e^2}{\hbar c}\right)^5\ln\frac{\hbar c}{e^2}\,.
\end{equation}
As follows from (\ref{51}) and (\ref{52}), this corresponds to 
\begin{equation}
\label{54}
\langle u^2\rangle=\frac{4r^{2}_{c}}{\pi}\frac{e^2}{\hbar c}\ln\frac{\hbar c}{e^2}\,,
\end{equation}
where $r_c=\hbar/mc\simeq 3.86\times 10^{-11}cm$ is the electron Compton length. 
\subsection{Electron-photon system}
\label{wellA}
Sometimes instead of the usual formalism of quantum electrodynamics, it is convenient to start with the multi-dimensional quantum mechanics of the electron-photon system. Photons can be treated 
as an infinite set of harmonic oscillators \cite{LANDAU}. In this method, proposed in Refs.~\cite{FEYN1,LEGG} and developed in further publications (see for example \cite{WEISS}), the Lagrangian 
of the total system 
\begin{eqnarray}
\nonumber
&&L=\frac{m}{2}(\dot{x}^2-\Omega^2x^2)+\frac{\mu}{2}\sum_k(|\dot{\xi_k}|^2-\omega^{2}_{k}|\xi_{k}|^2)\\
&&-x\sum_kc_k\xi_k -x^2\sum_k\frac{c^{2}_{k}}{2\mu\omega^{2}_{k}}
\label{101}
\end{eqnarray}
depends on ``photon'' coordinates $\xi_k$, where $\xi_{-k}=\xi^{*}_{k}$ and $\omega_k=ck$. The summation occurs on $-\infty<n<\infty$ with $k=2\pi n/L$ where $L$ is the system length. For 
simplicity we use one dimension as in Refs.~\cite{FEYN1,LEGG} and the harmonic potential $m\Omega^2x^2/2$ for the electron coordinate $x$. The cross-term in (\ref{101}) describes the 
``electron-photon'' interaction. The real coefficients $c_k=c_{-k}$ are specified below. The transition from classical description (\ref{101}) to quantum one is clear \cite{LEGG}. One should 
convert (\ref{101}) into the Hamiltonian with the substitution of the type $m\dot{x}\rightarrow -i\hbar\partial/\partial x$. 

The classical equations of motion follow from the Lagrangian (\ref{101})
\begin{equation}
\label{102}
m(\ddot{x}+\tilde{\Omega}^2 x)=-\sum_kc_k\xi_k,\hspace{0.5cm}\mu(\ddot{\xi}_k+\omega^{2}_{k}\xi_k)=-c_kx,
\end{equation}
where
\begin{equation}
\label{103}
\tilde{\Omega}^2=\Omega^2+\sum_k\frac{c^{2}_{k}}{2\mu\omega^{2}_{k}}.
\end{equation}
One can substitute the solution of the second equation (\ref{102}) into the first one. This results in the classical equation of motion \cite{LEGG}
\begin{eqnarray}
\label{104}
&&m\ddot{x}(t)+m\Omega^2 x\\
&&+\frac{2}{\pi}\int^{t}_{-\infty}dt_1\dot{x}(t_1)\int^{\infty}_{0}d\omega\,\eta(\omega)\cos\omega(t_1-t)=0,
\nonumber
\end{eqnarray}
where the summation rule and the viscosity coefficient are
\begin{equation}
\label{105}
\sum_k=\frac{L}{\pi c}\int^{\infty}_{0}d\omega,\hspace{0.5cm}\eta(\omega_k)=\frac{Lc^2(\omega_k)}{2c\mu\omega^{2}_{k}}.
\end{equation}
We use the notation $c(\omega_k)=c_k$. 
\subsection{Dissipative quantum mechanics}
\label{wellB}
The theory based on quantization of the Lagrangian (\ref{101}) is called in literature dissipative quantum mechanics. A variety of applications corresponds to the constant viscosity coefficient 
$\eta(\omega)=\eta_0$ \cite{LEGG}. In this case the classical equation of motion (\ref{104}) reads 
\begin{equation}
\label{106}
m\ddot{x}+\eta_0\dot{x}+m\Omega^2 x=0.
\end{equation}
The mean squared displacement is given by the fluctuation-dissipation theorem \cite{LEGG}
\begin{equation}
\label{107}
\langle x^2\rangle=\frac{i\hbar}{2\pi}\int^{\infty}_{-\infty}\cot\frac{\hbar\omega}{2T}\,\frac{d\omega}{m\omega^2-m\Omega^2+i\eta_0\omega}\,.
\end{equation}
At zero temperature $T=0$ and in the limit of small dissipation $\eta_0\ll m\Omega$
\begin{eqnarray}
\nonumber
&&\langle x^2\rangle=\frac{\hbar}{2m\Omega}\left[1-\frac{\eta_0}{\pi m}\int^{\infty}_{0}\frac{d\omega}{(\Omega+\omega)^2}\right]\\
&&=\frac{\hbar}{2m\Omega}\left(1-\frac{\eta_0}{\pi m\Omega}\right).
\label{108}
\end{eqnarray}
As follows from (\ref{108}), the particle becomes less fluctuative under dissipation. That is the fluctuation contraction compared to the free electron. In other words, coupling to the 
environment makes the particle more ``heavy''. This is generic with the known polaronic effect of enhancement of the effective electron mass due to the interaction with phonons in solids 
\cite{KITT}. In that case one can adjust $c_k$ in (\ref{101}) \cite{WEISS} where the set of oscillators describes phonons.

Also one can easily show (see for example \cite{IVLEV}) that the ground state energy of the system is 
\begin{equation}
\label{108a}
E_{tot}=\frac{\hbar\Omega}{2}+\frac{\hbar\eta_0}{2\pi m}\int^{\infty}_{0}\frac{d\omega}{\Omega+\omega}\,.
\end{equation}
The upper limit of the logarithmically divergent integral should be chosen using some additional arguments. For example, in the polaron problem $\omega$ should not exceed the Debye frequency.
\section{RELATION TO QUANTUM ELECTRODYNAMICS}
\label{qed}
Let us formally put 
\begin{equation}
\label{109}
\eta(\omega)=\frac{2e^2}{3c^3}\,\omega^2.
\end{equation}
In the classical limit we obtain from Eq.~(\ref{104}) three-dots-equation of the classical field theory
\begin{equation}
\label{110}
m\ddot{x}-\frac{2e^2}{3c^3}\,\dddot{x}+m\Omega^2 x=0,
\end{equation}
which is well discussed in textbooks, see for example \cite{LANDAU1}. 
\subsection{Mean squared displacement}
\label{qedA}
To calculate the mean squared displacement $\langle x^2\rangle$ we need the eigenfunctions $\psi^{(n)}(x)$ of the state $n$ of the harmonic oscillator with the mass $m$ and the frequency 
$\tilde{\Omega}$. We also need analogous functions $\varphi^{(0,1)}_{l}(\xi_l)$ for oscillators with $\mu$ and $\omega_l$. We define the functions
\begin{eqnarray}
\nonumber
&&|n;0\rangle=\psi^{(n)}(x)\prod_k\varphi^{(0)}_{k}(\xi_k)\\
&&|1;l\rangle=\psi^{(1)}(x)\varphi^{(1)}_{l}(\xi_l)\prod_{k\neq l}\varphi^{(0)}_{k}(\xi_k).
\label{111}
\end{eqnarray}
The first order correction, with respect to the interaction, to the total wave function of the system (\ref{101}) is \cite{LANDAU2}
\begin{equation}
\label{112}
\Psi_1=\sum_l\frac{c_l\langle1;l|x\xi_l|0;0\rangle}{(-\hbar\Omega-\hbar|\omega_l|)}\,|1;l\rangle.
\end{equation}
The second order correction has the form \cite{LANDAU2}
\begin{eqnarray}
\nonumber
&&\Psi_2=\sum_lc^{2}_{l}\frac{\langle 2;0|x\xi_l|1;l\rangle\langle1;l|x\xi_l|0;0\rangle}{(-\hbar\Omega-\hbar|\omega_l|)(-2\hbar\Omega)}\,|2;0\rangle\\
&&-\frac{1}{2}\,|0;0\rangle\sum_lc^{2}_{l}\frac{(\langle1;l|x\xi_l|0;0\rangle)^2}{(-\hbar\Omega-\hbar|\omega_l|)^2}.
\label{113}
\end{eqnarray}
The mean squared displacement is given by
\begin{equation}
\label{114}
\langle x^2\rangle=\int x^2dx\prod_kd\xi_k(\Psi_0+\Psi_1+\Psi_2)^2,
\end{equation}
where $\Psi_0=|0;0\rangle$. In the expression (\ref{114}) one should account for quadratic corrections $c^{2}_{l}$ only. The first term in (\ref{114}) produces the known contribution 
$\hbar/2m\tilde{\Omega}$ which should be expanded up to the second order according to the definition (\ref{103}). It is not difficult to collect all quadratic corrections in Eq.~(\ref{114}).
Omitting simple calculations we arrive to the fluctuation contraction 
\begin{equation}
\label{115}
\langle x^2\rangle=\frac{\hbar}{2m\Omega}\left[1-\frac{1}{\pi m}\int^{\infty}_{0}\frac{d\omega\eta(\omega)}{(\Omega+\omega)^2}\right].
\end{equation}

We see that the result (\ref{115}) coincides with (\ref{108}) if to put $\eta(\omega)$ to be constant. Since $\eta(\omega)\sim\omega^2$ the integral in (\ref{115}) is divergent at large frequencies.
In the polaron problem such integration should be cut off on the Debye frequency. In the interaction with photons there is no an artificial upper frequency. We do not consider limits of 
applicability of quantum electrodynamics. In that case, according to rules of quantum electrodynamics, an expression should be regularized by subtraction of a divergent part \cite{LANDAU}. This 
corresponds to the substitution in Eq.~(\ref{115})
\begin{equation}
\label{116}
\frac{1}{(\Omega+\omega)^2}\rightarrow\frac{1}{(\Omega+\omega)^2}-\frac{1}{\omega^2}.
\end{equation}
As a result, we obtain the fluctuation spreading compared to the free electron
\begin{equation}
\label{117}
\langle x^2\rangle=\frac{\hbar}{2m\Omega}+\frac{\hbar}{\pi m^2}\int^{\infty}_{0}\frac{d\omega\eta(\omega)}{\omega(\Omega+\omega)^2}\,.
\end{equation}
The integral in (\ref{117}) diverges at large frequencies only logarithmically. This divergence is not required a further regularization since it is related to the non-relativistic restriction
used in the model (\ref{101}). Due to that the integration in Eq.~(\ref{117}) is restricted by $mc^2/\hbar$ \cite{MIGDAL}. Finally, with the definition (\ref{109}),
\begin{equation}
\label{118}
\langle x^2\rangle=\frac{\hbar}{2m\Omega}+\frac{2r^{2}_{c}}{3\pi}\,\frac{e^2}{\hbar c}\ln\frac{mc^2}{\hbar\Omega}\,.
\end{equation}

In three dimensions the system (\ref{101}) is supplemented by two analogous ones containing $\{y,\eta_k\}$ and $\{z,\zeta_k\}$ in addition to $\{x,\xi_k\}$. By means of Eq.~(\ref{118}) we define 
the parameter ($u^2=x^2+y^2+z^2$)
\begin{equation}
\label{124}
r^{2}_{T}=\langle u^2\rangle=3\left(\langle x^2\rangle-\frac{\hbar}{2m\Omega}\right)=\frac{2r^{2}_{c}}{\pi}\,\frac{e^2}{\hbar c}\ln\frac{mc^2}{\hbar\Omega}\,,
\end{equation}
which is the mean squared amplitude of electron ``vibrations'' due to the interaction with photons in three dimensions. 

One can apply the result (\ref{124}) to hydrogen atom putting $\hbar\Omega\sim me^4/\hbar^2$ (Rydberg energy) \cite{MIGDAL}. This leads to the expression (\ref{54}) with the logarithmic accuracy. 
Since frequencies involved are not large, $\hbar\omega <mc^2$, the non-relativistic approach (\ref{101}) is applicable for calculations with that accuracy. To go beyond the logarithmic accuracy 
in (\ref{118}) the non-relativistic approach is not sufficient \cite{LANDAU}. 
\subsection{Energy of the state}
\label{qedB}
One can apply the usual perturbation theory to the multi-dimensional quantum mechanical system (\ref{101}) to calculate the energy correction \cite{LANDAU2}
\begin{equation}
\label{119}
E_{tot}-E=\langle x^2\rangle\sum_k\frac{c^{2}_{k}}{2\mu\omega^{2}_{k}}+\sum_lc^{2}_{l}\frac{\left(\langle 1;l|x\xi_l|0;0\rangle\right)^2}{(-\hbar\Omega-\hbar|\omega_l|)}\,.
\end{equation}
Substituting matrix elements for harmonic oscillator and accounting for the summation rule (\ref{105}), we arrive to 
\begin{equation}
\label{120}
E_{tot}=E+\frac{\hbar}{2\pi m}\int^{\infty}_{0}\frac{d\omega\eta(\omega)}{\Omega+\omega}.
\end{equation}
In the case of constant $\eta(\omega)=\eta_0$ the result (\ref{120}) coincides with (\ref{108a}).

In the divergent integral (\ref{120}), according to quantum electrodynamics, one should make the regularization 
\begin{equation}
\label{121}
\frac{1}{\Omega+\omega}\rightarrow\frac{1}{\Omega+\omega}-\frac{1}{\omega}+\frac{\Omega}{\omega^2},
\end{equation}
which is similar to (\ref{116}). The result is
\begin{equation}
\label{122}
E_{tot}=E+\frac{\hbar\Omega^2}{2\pi m}\int^{\infty}_{0}\frac{d\omega\eta(\omega)}{\omega^2(\Omega+\omega)}.
\end{equation}
According to (\ref{117}), with the logarithmic accuracy the result (\ref{122}) can be written as 
\begin{equation}
\label{123}
E_{tot}=E+\frac{m\Omega^2}{2}\left(\langle x^2\rangle-\frac{\hbar}{2m\Omega}\right),
\end{equation}
where $\langle x^2\rangle$ is determined by Eq.~(\ref{118}). The part in parenthesis relates to electron-photon fluctuations. In the case of one-dimensional harmonic oscillator the energy 
(\ref{123}) corresponds to the Lamb shift and is analogous to (\ref{52}). 
\subsection{Comments}
\label{qedC}
Above we use the multi-dimensional non-relativistic quantum mechanics, with continuum variables, related to the Lagrangian (\ref{101}). Coupling to the environment is chosen in a way to get in
the classical limit the famous three-dots-equation (\ref{110}). In frameworks of this quantum mechanical approach the mean squared displacement of the electron is reduced by photons (\ref{115})
(fluctuation contraction). In other words, the electron becomes more ``heavy''. Accordingly the total energy (\ref{120}) is increased producing that ``heaviness''. This is generic with the polaron 
effect when the effective electron mass increases due to the interaction with phonons. Divergent integrals should be cut off by the Debye frequency. This situation corresponds to dissipative 
quantum mechanics. 

In contrast to dissipative quantum mechanics, under the interaction with photons there is no a cut off frequency. We do not consider limits of applicability of quantum electrodynamics. This is a 
situation of quantum electrodynamics when one should regularize integrals. It means a subtraction of divergent parts until a result becomes convergent \cite{LANDAU}. 

That type of regularization is performed in our case in Secs.~\ref{qedA} and \ref{qedB} which leads to convergent results. From the standpoint of the quantum mechanical system (\ref{101}) the
regularization is a formal subtraction of some extra terms. As a result, the electron becomes less ``heavy'' (fluctuation spreading (\ref{117})) and accordingly the total energy is reduced. In 
other words, this energy reduction naturally assists fluctuation spreading. The regularization done above results in the correct (with the logarithmic accuracy) Lamb shift following from quantum 
electrodynamics. This relates to the both Coulomb field and harmonic potential. 

It is amazing that despite the model (\ref{101}) differs from quantum electrodynamics in a few aspects, it results in three-dots-equation and the correct Lamb shift. One can say in the different 
way, as soon as we fixed condition for validity of three-dots-equation this results in the correct Lamb shift. 

An applicability of the non-relativistic approach (\ref{101}) is due to small frequencies $\hbar\omega<mc^2$ involved into (\ref{117}) and (\ref{122}) \cite{MIGDAL}. Under that condition 
three-dots-equation (\ref{110}) is also applicable. Calculations beyond the logarithmic accuracy requires the non-relativistic region. 

Below we need details of the mechanism of electron mass generation which is hidden in the regularization.
\section{ANOMALOUS ELECTRON STATES}
\label{anom} 
Let us switch off the electron-photon interaction. If an energy in the single electron wave equation does not coincide with an eigenvalue the proper wave function becomes singular. In Appendix the 
linear singularity along the $z$ axis is considered. The kinetic energy term $-\hbar^2\nabla^2/2m$ in Eq.~(\ref{A3}) is singular as $\delta(\vec r)$. This term is formally compensated in (\ref{A3}) 
by the artificial $\delta(\vec r)$ term playing a role of a potential well. Such counter-term does not exist in reality and therefore the singular state is not physical. The obvious question is 
that what appears to this singular state if you switch on the electron-photon interaction. 
\subsection{Hypothesis of anomalous states}
\label{anomA}
Under the electron-photon interaction the electron gets more spread in space according to (\ref{124}). The singular line participates in ``vibrations'' resulting in smearing of the singularity
in the kinetic energy in (\ref{A3}). To get the resulting state physical ({\it anomalous electron state}) the counter-term, compensating distributed singularities, should exist. Let us propose a 
hypothesis that such term exists. In this case that term reminds a smeared $\delta(\vec r)$ like a narrow potential well. 

It is convenient to consider formation of anomalous state step by step. Suppose that without the interaction with photons the wave function has the above singularity along the $z$ axis. Let us 
switch on the interaction with photons implying wave vectors in (\ref{101}) to be restricted by some large value $k_{max}=\omega_{max}/c$. 

First, we consider pure multi-dimensional quantum mechanical problem (\ref{101}). Far away from the $z$ axis the state is hardly violated by the interaction with photons. One can track this
exact stationary solution, with the total energy $E_{tot}$, in multi-dimensional space from large to small $r$. This is possible due to locality of the system described by differential equations. 
In this process the state remains singular at small $r$. We consider the case when the phase of the wave function is not changed after going around the $z$ axis. The state, continued from the 
infinity, comes to the singularity at $\vec r=\vec u(z)$
\begin{equation}
\label{123a}
\psi\sim\frac{1}{|\vec r-\vec u|}\,,
\end{equation}
where $\psi$ is the bispinor related to the electron (compare with (\ref{A5})). Without the electron-photon interaction $\vec u(z)=0$. A narrow distribution of singularity positions $\vec u(z)$ 
depends on a choice of photon degrees of freedom. 

After the average on photons $\langle\vec u\rangle=0$ and the state is still non-physical, as a superposition of singular states, with the contracted fluctuations analogous to (\ref{115}) and 
the enhanced energy analogous to (\ref{120}), where $\omega<\omega_{max}$. 

Second, let us perform the regularization. In this process the electron density gets spread into the thread of the radius $r_T$ around the former singularity line. This is possible only if our
hypothesis is valid: the energy should locally reduces inside the thread providing a counter-term to support the extra electron density on the thread. After that one can remove the $\omega_{max}$ 
restriction and the state becomes physical. In principle, that energy reduction is not contradictory. From the stand point of multi-dimensional quantum mechanics regularization corresponds to a 
formal subtraction of some extra terms. 

So the hypothesis of anomalous electron states is formulated below. The initial state (with the singularity along the $z$ axis), if you switch on the interaction with photons, gets smeared  within 
the thread directed along the $z$ axis. The thread radius is small but finite and therefore the resulting anomalous electron state is not singular. 

One can consider an one-electron quantum mechanical problem with given space-time dependence of electromagnetic four-potentials $A_{\mu}(t,\vec R)$. In this case the electron wave function is of 
the same type as (\ref{123a}). The electron density
\begin{equation}
\label{123b}
n\{A_{\mu}(t,\vec R)\}\sim\frac{r^{2}_{c}}{\left[\vec r-\vec u\{A_{\mu}(t,\vec R)\}\right]^2}
\end{equation}
depends on non-stationary singularity positions $\vec u$ localized close to the $z$ axis and determined by electromagnetic four-potentials $A_{\mu}$. The form (\ref{123b}) is valid at the 
``adiabatic'' condition when photon frequencies $\omega<c/|\vec r-\vec u|$.

The justification of the hypothesis of anomalous states is given in Sec.~\ref{anomB}.
\subsection{Relation to the generation of electron mass}
\label{anomB}
In quantum electrodynamics a direct calculation of the mass correction, caused by the interaction with photons, results in the part logarithmically divergent at large momenta \cite{FEYN,LANDAU}. 
The regularization of the mass expression leads to the reduction of the infinite electromagnetic mass down to the physical value \cite{LANDAU}. Regularization parameters are adjusted to the 
physical mass and formed within the electron radius. The regularization process hides any physical mass-generating mechanism.

In the Standard Model masses of electron, other leptons, $W^{\pm}$ and $Z$ weak bosons, and quarks are generated by Higgs mechanism which involves the scalar Higgs field \cite{HIG,ENG,GUR}. 
Electron, as a fermion, acquires its mass by the connection between the fermion field $\psi$ and the Higgs field $\phi$. The Lagrangian 
\begin{equation}
\label{126}
L=\bar{\psi}i\gamma^{\mu}\tilde{D}_{\mu}\psi-G\bar{\psi}\phi\psi+L_{H}(\phi)+L_g
\end{equation}
contains the Higgs part
\begin{equation}
\label{126a}
L_{H}(\phi)=(D_{\mu}\phi)^+D^{\mu}\phi+\mu^2\phi^+\phi-\lambda(\phi^+\phi)^2
\end{equation}
and the gauge part $L_g$ that, for pure electromagnetic field, would be $-F^{\mu\nu}F_{\mu\nu}/4$ where $F_{\mu\nu}=\partial_{\mu}A_{\nu}-\partial_{\nu}A_{\mu}$. The Yukawa term, depending on the 
coupling $G$, is written in (\ref{126}) in a schematic form. The covariant derivatives $\tilde{D}_{\mu}$ and $D_{\mu}$ contain, in addition to partial derivatives $\partial_{\mu}$, the parts
depending on gauge fields $W^{\pm}_{\mu}$, $Z_{\mu}$, and $A_{\mu}$. In (\ref{126}) $\gamma^{\mu}$ are the Dirac matrices and we use the units with $c=\hbar=1$.

The physical electron mass appears due to a finite expectation value $|\langle\phi\rangle|$ (in the Yukawa term) that relates to the ground state of $L_H$ \cite{HIG,ENG,GUR}. 

The above formalism of the Standard Model can be applied to the case when the electron density is singular (\ref{123b}). Not too close to the singularity fluctuations of fields 
$\phi$, $W^{\pm}_{\mu}$, and $Z_{\mu}$ are not significant and the electron density (averaged on the fermion field) is given by Eq.~(\ref{123b}). The obvious question is that what happens to the 
density (\ref{123b}) closer to the singularity. As the field $A_{\mu}$, one can consider the fields $W^{\pm}_{\mu}$ and $Z_{\mu}$ also as macroscopic ones taking in mind a further average on these 
fields. Then closer to the singularity its position $\vec u\{W^{\pm}_{\mu},Z_{\mu},A_{\mu}\}$ depends on the electroweak set.

As follows from (\ref{126}), 
\begin{equation}
\label{127}
\langle[-D^{\mu}D_{\mu}\phi^++\mu^2\phi^+-2\lambda\phi^+(\phi^+\phi)]\rangle=\langle G\bar{\psi}\psi\rangle\,,
\end{equation}
where angle brackets mean the average on fields $\psi$ and $\phi$ at given space-time dependence of gauge fields $W^{\pm}_{\mu}$, $Z_{\mu}$, and $A_{\mu}$. Usually the expectation value 
$\langle\phi\rangle$ is slightly influenced by the right-hand side in (\ref{127}) that is proportional to $mc^2/100GeV\sim 10^{-5}$. In the singularity case the right-hand side of (\ref{127}) 
is generic with the electron density (\ref{123b}). The singular $n$ increases at $r\sim 10^{-13}cm$ by five orders of magnitude compared to its usual value (at $r_c$). According to (\ref{127}), 
the singularity of its right-hand side produces a singularity in expectation values related to $\langle\phi\rangle$. Therefore processes, similar to the electron mass generation, enter the game 
closer to the singularity. 

When $\langle\phi\rangle$ increases approaching the singularity, the squared gradient term in (\ref{127}) becomes large. Therefore terms with higher power of gradients, omitted in (\ref{127}) in
the usual case, now should be taken into account. Corresponding terms in the Lagrangian can be, for example, even powers of $(D_{\mu}\phi)^+D^{\mu}\phi$ that does not violate the symmetry of the 
problem. In Eq.~(\ref{127}) those terms are perturbations not too close to the singularity. Under this condition $\langle\phi\rangle$ acquires the series of terms proportional to 
$1/(\vec r-\vec u)^{2n}$ where each next term is smaller than previous one. According to (\ref{126}), a series of the same type also appears in the electron density $n$. Approaching the singularity, 
those terms become large and the resulting singularity can be established from general scaling arguments. As follows from (\ref{126}), $\langle\phi\rangle$ should scale as $1/|\vec r-\vec u|$. In
principle, that limiting position of the singularity can be complex. Then the relation (\ref{127}), supplemented by higher derivatives, specifies a type of singularity of $n$. 

In this scheme the increase of the electron kinetic energy term, under approaching the singularity, is automatically compensated by the counter-term related to the enhancement of 
$\langle\phi\rangle$. 

Then the final step has to be done. The solution, taken at given gauge fields $W^{\pm}_{\mu}$, $Z_{\mu}$, and $A_{\mu}$, should be averaged on these degrees of freedom. Analogous average on 
$A_{\mu}$ is performed in quantum electrodynamics. The average on large mass fields $W^{\pm}_{\mu}$ and $Z_{\mu}$ results in short range fluctuations. These fluctuations wash out the singularity 
on the distances that can be identified with the electron radius. The further average, on the field $A_{\mu}$, additionally smears out this distribution on larger scale $r_T$. The radius $r_T$ is 
determined not by mass-generating phenomena but by photons with frequencies $\hbar\omega<mc^2$ (Sec.~\ref{qedA}). 

In the initial form (\ref{123b}), that is not too close to the singularity, the electron density is localized in the thing thread along the $z$ axis. Therefore the smeared $n$ is distributed also 
within the thread. As a result, the narrow but smooth peak of the electron density $n$ along the thread is accompanied by the narrow and smooth counter-term. 

The formation of the counter-term is an essential point. This term, analyzed above, ultimately originates from a local reduction of the vacuum energy of the mass-generating field. This anomalous 
vacuum looks as ``well'' localized along the thread which is placed at a minimum of an external potential. Without the external potential (for a free electron) $r_T=\infty$ and anomalous effects 
disappear. 

There is the famous example of creation of the well by spatial variations of the vacuum energy. This is the Casimir effect when the zero-point photon energy $\sum\hbar\omega/2$ becomes variable 
due to a spatial variation of the photon density of states \cite{LANDAU,CAS}.

So the hypothesis of anomalous electron states, proposed in Sec.~\ref{anomA}, is supported by the above arguments referred to the mass-generating mechanism.

One can put a question about anomalous vacuum related to quarks. Their mass generation and mixing are also due to the Higgs mechanism with the assistance of Yukawa terms.
\subsection{Features of anomalous states}
\label{anomC}
With the electron-photon interaction the singularity of $F(\vec r\,)$ is distributed as roughly $\langle F(\vec r-\vec u)\rangle$ and the kinetic part in (\ref{A3}) can be approximately estimated 
as 
\begin{equation}
\label{125}
-\frac{\hbar^2}{2m}\nabla^2\big\langle F(\vec r-\vec u)\big\rangle\sim\frac{\hbar^2}{mr^{2}_{T}}F(r\sim r_T). 
\end{equation}
This part is localized at $r\lesssim r_T$ and is supported by the counter-term (Sec.~\ref{anomB}). This term can be interpreted as $\delta(\vec r)$ which is smeared out turning to the certain 
function of the amplitude $1/r^{2}_{T}$ and localized at $r\lesssim r_T$. This can be called anomalous well. 

The singularity along the $z$ axis turns to the subatomically narrow thread of the small but finite radius $r_T\sim 10^{-11}cm$. Within this thread the enhanced electron energy 
$(m^2c^4+\hbar^2 c^2/r^{2}_{T})^{1/2}\simeq \hbar c/r_T\sim 1MeV$ coexists with the reduction of the energy $\hbar c/r_T$ responsible for anomalous well. 

Without the interaction with photons in the potential $m\Omega^2R^2/2$ the wave function, singular on the line, exist at any energy $E_{tot}$ below the usual Lamb shifted value. Under the 
interaction with photons this state goes over into anomalous one regardless of $E_{tot}$. Therefore the energy spectrum of anomalous states is continuous. This can be interpreted as formation of 
the narrow ($\sim 10^{-11}cm$) and deep ($\sim 1MeV$) anomalous well along the $z$ axis which is adjustable to an electron state. Before we consider the harmonic potential. The results obtained 
also relate to a more general potentials.  

Anomalous electron states are exact and non-perturbative. Indeed, the electron density depends on $(\vec r-\vec u)$, where the both displacements are of the same order at $r<r_T$. The depth 
$\hbar c/r_T$ of anomalous well, formed by the reduction of the vacuum energy, is estimated as 
\begin{equation}
\label{9}
{\rm well\,\,depth}\sim mc^2\sqrt{\frac{\hbar c}{e^2}} 
\end{equation}
and cannot be obtained by the perturbation theory on $e^2/\hbar c$. More precise estimate gives for the energy (\ref{9}) a few $MeV$. 

Since the states are exact, their continuous spectrum is not decaying, that is ${\rm Im}\,E_{tot}=0$. The continuous non-decaying spectrum of a particle in a potential well is not forbidden in 
nature. Such spectrum is revealed in Ref.~\cite{IVLEV} on the basis of the exact solution. 

There is the qualitative explanation why anomalous states are non-decaying (as states in \cite{IVLEV}). The narrow region $r_T$ plays a role of the
point where the electron is tightly connected to electromagnetic coordinates and is dragged by them. One can treat the electron to be localized in 
that region. Under photon emission the narrow region would oscillate increasing the electron kinetic energy. This prevents the electron to lose its 
total energy and therefore results in non-decaying states.
\section{LINK TO THE EXPERIMENTS}
\label{exp}
In this section we analyze how anomalous states relate to the experimental observations \cite{IG10,IG11}. 
\subsection{Creation of anomalous states}
\label{expB}
Creation of $MeV$ depth anomalous well is energetically favorable. But such well can be formed at points where the electron is in a potential minimum. To create that minimum a few conduction 
electrons should be localized near lattice sites to dominate their positive charges. This is possible since the resulting energy gain is large (of the $MeV$ scale). The thread of the anomalous
state can be not necessary a straight line but of various forms including rings. 

The usual state of a conduction electron and the anomalous one are different eigenstates of the total system. A perturbation, which transfers one state into another should be of a short range 
in space. Otherwise the transition matrix element would be small due to the difference in spatial scales (the Bohr radius and $r_T\sim 10^{-11}cm$) of two states. The optimal spatial scale of a
perturbation is $r_T$. 

The charge density, varying is space on the typical distance $r_T$, can be created by an incident charged particle which is reflected by lattice sites of the solid. The resulting density, related
to such particle, is due to interference of its incident and reflected waves. This charge density is approximately proportional to $\cos(2R\sqrt{2M_pE_p}/\hbar)$ where $M_p$ is the particle mass 
and $E_p$ is its energy. For example, for deuterons $M_p\simeq 3.346\times 10^{-24}g$ one can estimate
\begin{equation}
\label{54a}
{\rm charge\,\,density}\sim\cos\left[1.96\frac{R}{r_T}\sqrt{E_p(keV)}\right],
\end{equation}
where $r_T$ is taken to be $10^{-11}cm$. As a result, thread segments can be formed with the following association into rings or lines connected two lattice sights to reduce the energy. 

Also a direct absorption of the quantum $E_p$ is possible with the transition of the electron down to $\sim 1keV$ in the anomalous well. We see that one can irradiate the surface of the solid by 
ions with the energy of approximately $1keV$ to produce anomalous electron binding.
\subsection{X-ray emission}
\label{expC}
The goal of this paper is to reveal a mechanism which is compatible with the four unusual conditions of Sec.~\ref{karD}. 

The statement about infinite lifetime of states in anomalous well is referred to a static potential whose minimum has the fixed position in space. In a crystal lattice the minimum position 
is determined by positions of lattice sites. These sites thermally vibrate since experiments \cite{IG10,IG11} were conducted at room temperature which is on the order of the Debye energy
$\hbar\omega_D$. Therefore the position of the anomalous state, containing electron, also vibrates. This result in emission of electromagnetic waves (Bremsstrahlung) and hence to the appearance 
of a finite lifetime of states in the anomalous well. The rate of the energy emitted \cite {LANDAU1} is opposite in sign to 
\begin{equation}
\label{55}
\frac{dE_{tot}}{dt}=-\frac{2e^2}{3c^3}\,\dot{\vec v}\,{^2},
\end{equation}
where $\vec v$ is the velocity of electron thermal vibrations. One can approximate $\vec v\,{^2}\sim \omega^{2}_{D}(\hbar/M\omega_D)$, where $M$ is the mass of the lattice site. By means of that
\begin{equation}
\label{56}
\frac{dE_{tot}}{dt}=-\frac{e^2}{\hbar c}\left(\frac{\hbar\omega_D}{Mc^2}\right)\omega_D(\hbar\omega_D)\sim -10^{-14}\omega_D(\hbar\omega_D).
\end{equation}
The electron in the anomalous well steady goes down in energy according to (\ref{56}). So states in the anomalous well are of the continuous spectrum and of long but finite lifetime. This is the 
basis for laser generation with the continuous spectrum observed in experiments \cite{IG10,IG11}. 

In the usual three energy level scheme of laser operation the intermediate level should be long-living to create a population inversion. The subsequent photon emission is monochromatic. In our case 
the entire continuous spectrum is of long-living states. These conditions also provide a laser emission but of the continuous spectrum. The continuous energy spectrum of emitted collimated X-rays is 
in the range of $keV$ as follows from Sec.~\ref{kar}. This reminds a continuous set of monochromatic lasers.

Since anomalous well is deep ($\sim 1MeV$) it is favorable to accept other electrons to the well during irradiation of the sample by ions. The maximum number $N_{max}$ of electrons in an anomalous well 
can be estimated from the condition $N^{2}_{max}e^2/r_T\sim 1MeV$. This gives $N_{max}\sim 10$. 

Another unusual phenomenon, observed in \cite{IG10,IG11}, is an emission of collimated bursts after switching off the irradiation by external ions. This post-irradiation emission, in the form of 
separate bursts, was observed during 20 hours. After the irradiation, resting electrons close to well bottoms serve as a short scale perturbation (analogous to charge density produced by 
irradiating ions) for conduction electrons to get them converted into anomalous states. This can be the mechanism of generation of post-irradiation bursts. 

The energy relaxation in $MeV$ depth wells may be accompanied, besides $keV$ bursts emission, also by the emission of quanta in the $MeV$ range. Those high-energy quanta can cause nuclear 
transmutations in lattice sites. Multiple nuclear transmutations were reported in Refs.~\cite{IG10,IG11}. 

We see that the proposed phenomenon of anomalous electron states is compatible with the four conditions outlined in Sec.~\ref{kar}D.

The energy of X-ray bursts ($keV$ range) and other quanta ($MeV$ range) is generated when the electron goes down in energy in the anomalous well. This well is created by a variation in space of
the mass-generating field (Sec.~\ref{anomB}). The emitted energy is stored before each pulse but not 20 hours back before switching off the ion irradiation. Under a steady irradiation by $keV$ 
ions a steady energy conversion into $MeV$ quanta occurs. A source of this energy is the vacuum of the mass-generated field. 
\section{DISCUSSIONS}
\label{disc}
The paradoxical observations \cite{IG10,IG11} stay apart from a variety of effects caused by irradiation of solids. The essential point is that experiments \cite{IG10,IG11} were repeatedly 
performed for years and could be reproduced any time on demand. Indeed, the array of macroscopic laser bursts unlikely is an artifact. The phenomena \cite{IG10,IG11} cannot be explained by 
mechanisms which usually work in the field. The extraordinary features, including an appearance of unexpected $MeV$ energies in condensed matter physics, require accounting for different mechanisms. 
The rigorous conclusion about this mechanism is a substantially subatomic nature of it. Atomic phenomena occur at the Bohr radius and correspond to $eV$ but not $keV$ or $MeV$ energies. 

Nuclear processes, as an example of subatomic ones, are not responsible for the effects observed. The $keV$ energies of irradiating ions are not sufficient, as known, to directly cause nuclear 
processes due to high Coulomb barriers around nuclei. Also the existence of long-living states in the $keV$ range is incompatible with nuclear spectroscopy. This points to processes where not
nuclei but electrons are substantially involved. In this case the electron kinetic energy of the $MeV$ range implies the scale of $10^{-11}cm$ of spatial localization of an electron according to 
the uncertainty principle. 

On the other hand, electrons, interacting solely with crystal lattice, including the static field and phonons, cannot relate to $MeV$ energies. It is impossible to get $10^{8}$ phonons (of the
energy $10^{-2}eV$ each) coherently converted into the $MeV$ energy in the crystal providing a sudden acoustic shock. Also electrons, interacting with photons only (no mean-field forces on 
electrons), cannot lead to pronounced subatomic phenomena. In this case there is the usual Lehmann representation of electron propagator excluding such effects (without those forces $r_T$ is 
infinite).

We see that solely the combination (electron)-(photons)-(crystal lattice) may underlie the subatomic mechanism resulting in the four experimental features focused on in Sec.~\ref{karD}. But
the link between those issues is not obvious. In this paper, to build up such a link, the concept of anomalous electron states is proposed which consists of the following parts.

(1) The electron wave function which, in the absence of interaction with photons, is singular on a line. In this case the kinetic energy term $-\hbar^2\nabla^2/2m$ is singular on that line. This 
singularity is not physical since it is not compensated by a singular potential well in the wave equation. 

(2) An addition to the wave equation of electromagnetic four-potentials, with given space-time dependence, shifts singularity positions from the straight line. This singularity line goes over into 
a curve in the vicinity of that line.

(3) According to the Standard Model, the singularity in space of the electron density results in a singularity of the expectation value of the Higgs field $\langle\phi\rangle$. The increase of the
kinetic energy term, under approaching the singularity, is automatically compensated by the enhancement of $\langle\phi\rangle$. 

(4) The solution, taken at given space-time dependence of electroweak fields, should be averaged on all values of these fields. The resulting electron density becomes smooth and localized within 
the thread of the small radius along the initial line. This anomalous state is physical. 

It is unusual that the phenomenon, recalling pure condensed matter one, involves physics of the Higgs field. 

The resulting electron anomalous state contains the thin thread, of the radius $r_T\sim 10^{-11}cm$, where the deep ($\hbar c/r_T\sim 1MeV$) anomalous well is localized. The origin of this ``well''
is due to a local reduction of the vacuum energy of the Higgs field. The appearance of the spatial scale $r_T\sim 10^{-11}cm$ is not surprising. This distance is involved even into the usual 
phenomenon of the Lamb shift when the wave function is smooth on the distance $r_T$. In this case the ''vibrating'' electron probes various parts of the static potential and therefore changes its 
energy a little. 

Anomalous electron state cannot exist in vacuum. This negatively charged state should be localized at the minimum of some mean-field potential. Such equilibrium points can exist in a metal due 
to the action of crystal sites and redistributed conduction electrons. This redistribution is energetically favorable since it leads to the high energy gain in anomalous well.

Transitions down in energy in anomalous $MeV$ wells produce $MeV$ quanta which can cause nuclear transmutation observed in \cite{IG10,IG11}. We emphasize that those processes, involving the $MeV$ 
energy range, are not due to nuclear fusion, as supposed in Refs.~\cite{IG10,IG11}, but of the electron origin. 

Another issue is the emission of X-ray laser pulses by various metals after switching off the external irradiation by $keV$ ions 20 hours back (Sec.~\ref{kar}). The usual purse quarter, irradiated 
yesterday evening by $keV$ ions, can emit X-ray laser bursts. The theory, proposed in this paper, points to the mechanism of this post-irradiation emission (Sec.~\ref{exp}). Detailed studies of 
the phenomenon, such as duration of emitted bursts, formation of the laser emission, its exact frequency range, etc., are outside of frameworks of this paper.  

The emitted energy comes from the vacuum of mass-generating field. Further development of X-ray lasers, which operate consuming vacuum energy, is promising. The steady conversion of $keV$ energy of 
irradiating ions into $MeV$ energy of generated quanta (steady extraction of vacuum energy) can be significant for applications. 

Anomalous electron states are responsible not only for phenomena observed in experiments \cite{IG10,IG11}. The recent observations of anomalous oscillations of magnetoresistance in superconductors 
\cite{MIL} provide another mysterious example generic with \cite{IG10,IG11}. The paradoxical universality of the oscillation phenomenon (in particular, material independence) can be explained 
solely by a subatomic mechanism. The proposed subatomic mechanism, based on electron anomalous states \cite{IVLEV1}, provides an excellent qualitative explanation of the experimental results
\cite{MIL}. Experimental studies \cite{IG10,IG11,MIL} seem to initiate a different field of research. 

One can put a question about anomalous vacuum related to quarks. Their mass generation and mixing are also due to the Higgs mechanism with the assistance of Yukawa terms.
\section{CONCLUSIONS}
There is the link from the experiments \cite{IG10,IG11} to anomalous electron states formed in the metal. These states are of the subatomic size $10^{-11}cm$ and related to anomalous well of the 
approximate depth $1MeV$. Such high energy phenomena are unusual in condensed matter physics. Anomalous vacuum of the mass-generating field is involved which is related to anomalous well with 
the depth proportional to $\sqrt{\hbar c/e^2}$. The electron moves in anomalous well down in energy resulting in the unusual electromagnetic emission. 

\acknowledgments
I thank M. Kunchur, J. Knight, and N. Nikolaev for discussions and remarks. This work was supported by CONACYT through grant number~237439.

\appendix*
\section{Singular solution of Dirac equations}
The static Schr\"{o}dinger equation formally has the solution which behaves as $\ln r$ at small $r$. We use cylindrical coordinates $r^2=x^2+y^2$. In that case the phase of the wave function is 
not changed after going around the $z$ axis. Below we establish the continuation of this singular solution to the region $r<r_c$ where one should use the Dirac formalism. In this case the wave 
function is the bispinor consisting of two spinors $\varphi$ and $\chi$ \cite{LANDAU}. Since we are interested by the singular wave function (large kinetic energy part) one can ignore, as the 
first step, the potential energy and consider free electron Dirac equations
\begin{eqnarray}
\label{A1}
&&(\varepsilon+i\hbar c\vec\sigma\nabla)\varphi-mc^2\chi=-\hbar^2c^2\Phi_0\delta(\vec r)\\
\nonumber
&&(\varepsilon-i\hbar c\vec\sigma\nabla)\chi-mc^2\varphi=-\hbar^2c^2\Phi_0\delta(\vec r).
\end{eqnarray}
Here $\varepsilon$ is the total relativistic energy, $\vec\sigma$ are Pauli matrices, and $\Phi_0$ is the certain constant spinor. We consider two-dimensional case when $z$ derivatives are zero. 
The solution of Eqs.~(\ref{A1}) is
\begin{eqnarray}
\label{A2}
&&\varphi=(\varepsilon+mc^2-i\hbar c\vec\sigma\nabla)F(\vec r)\\
\nonumber
&&\chi=(\varepsilon+mc^2+i\hbar c\vec\sigma\nabla)F(\vec r),
\end{eqnarray}
where one accounts for the relation $(\vec\sigma\nabla)(\vec\sigma\nabla)=\nabla^2$ and the spinor function $F(\vec r)$ satisfies the wave equation \cite{LANDAU}
\begin{equation}
\label{A3} 
-\nabla^2F -\Phi_0\delta(\vec r)=\frac{\varepsilon^2-m^2c^4}{\hbar^2c^2}F
\end{equation}
The solution of (\ref{A3}) is the Neumann function \cite{GRAD}
\begin{equation}
\label{A4} 
F(\vec r)=-\frac{\Phi_0}{4}N_0\left(\frac{r}{\hbar c}\sqrt{\varepsilon^2-m^2c^4}\,\right)
\end{equation}
with the asymptotics $N_0(z)\simeq (2/\pi)\ln z$ at small argument. Accordingly, at short distances two spinors are
\begin{equation}
\label{A5}
\varphi(r)=\left(\frac{\varepsilon+mc^2}{2\pi}\ln\frac{1}{r}+\frac{i\hbar c}{2\pi r^2}\,\vec\sigma\vec r\right)\Phi_0
\end{equation}
\begin{equation}
\nonumber
\chi(r)=\left(\frac{\varepsilon+mc^2}{2\pi}\ln\frac{1}{r}-\frac{i\hbar c}{2\pi r^2}\,\vec\sigma\vec r\right)\Phi_0
\end{equation}
In the standard representation $\Phi=\varphi+\chi$ and $\Theta=\varphi-\chi$
\begin{equation}
\label{A6} 
\Phi(\vec r)=\frac{\varepsilon+mc^2}{\pi}\,\Phi_0\ln\frac{1}{r},\hspace{0.5cm}\Theta(\vec r)=\frac{i\hbar c\Phi_0}{\pi r^2}\,\vec\sigma\vec r.
\end{equation}
At distances $r_c<r$ (non-relativistic limit) $\Theta$ is small compared to $\Phi$ and the wave function is the usual spinor $\Phi$.

\end{document}